# Schottky Barrier Heights of Defect-free Metal/ZnO, CdO, MgO and SrO Interfaces

Jiaqi Chen[1], Zhaofu Zhang[1], Yuzheng Guo[2], and John Robertson[1]

[1]*Department of Engineering, Cambridge University, Cambridge, CB2 1PZ, United Kingdom*
[2]*College of Engineering, Swansea University, Swansea, SA1 8EN, United Kingdom*

The Schottky barrier heights (SBHs) of defect-free interfaces of ZnO, CdO, MgO and SrO with various metals and different terminations are investigated by density functional supercell calculations. The oxide bands are corrected for their density functional band gap error by applying U-type treatment to their metal-d and O-p states where necessary. The p-type SBHs are found to decrease linearly with increasing metal work function. The pinning factor S of the non-polar and polar interfaces are similar for each oxide. S is found to be 0.26, 0.56, 0.74 and 0.96 for CdO, ZnO, MgO and SrO, respectively, with S increasing for increased oxide ionicity. The calculated pinning factors are generally consistent with the metal-induced gap states (MIGS) model in terms of variation with ionicity and dielectric constant. A significant shift of SBHs from the non-polar to the polar interfaces of 0.4 eV, 1 eV and 0.5 eV for ZnO, MgO and SrO, respectively, can be explained by an interfacial dipole. Our results are also useful to describe Co,Fe|MgO interfaces in magnetic tunnel junctions.

## I. Introduction

Metal-oxide contacts have attracted significant interest in numerous studies. It is important for applications including light emission [1,2], catalysis [3,4], field-effect transistors [5] and magnetic tunnel junctions [6]. One of the major concerns of the oxide's use in modern microelectronic devices is the contact resistance, which could be tuned by the interface Schottky barrier heights (SBH). Thus, understanding the formation mechanism of the SBH is very important.

The magnitude of SBH is traditionally understood by finding its dependence on the metal work function [7]. Weakly interacting surfaces follow the Schottky-Mott rule where the SBH is the difference between the semiconductor's electron affinity and the metal's work function. For stronger interactions, charge transfer between the metal Fermi level and states induced in the semiconductor band gap gives rise to a Fermi level pinning effect. The SBH can then be expressed as [8]:

$$\phi_n = S(\phi_M - \phi_S) + (\phi_S - \chi_S) \quad (1)$$

where $\phi_M$ is the metal work function, $\phi_S$ is the semiconductor's reference energy such as its charge neutrality level (CNL) and $\chi_S$ is the semiconductor's electron affinity. The degree of pinning is defined by a pinning factor S as deduced by Cowley and Sze [9],

$$S = \frac{1}{1 + \frac{e^2 N \delta}{\varepsilon \varepsilon_0}} \quad (2)$$

where N is the areal density of interface states, δ is their decay length into the oxide and $\varepsilon$ is the dielectric constant of the semiconductor interface region. The S factor varies between 0 (for a strongly pinned interface, i.e., Bardeen limit) and 1 (for no pinning interface, i.e., Schottky limit).

Various models have been proposed for the underlying pinning mechanism and the parameters which determine the value of the SBH. Presently, the magnitude of SBH is thought to depend on three main factors: the intrinsic metal-induced gap states (MIGS), interfacial dipole layers and any extrinsic interfacial point defects [8,10]. For semiconductors of moderate band gap, the most important factor controlling S is generally the MIGS [8,10–13]. This is because the tail of metal wave functions decaying into the semiconductor gap is the main cause of pinning. In this case, Mönch [10,13] found that S followed the empirical formula:

$$S = \frac{1}{1+0.1(\varepsilon_\infty-1)^2} \qquad (3)$$

However, as the band gap widens, the extent of the MIGS reduces and their pinning effect declines. Defect states then tend to become more important and dominate any pinning effect.

In addition, as ionicity increases, dipole layers can arise at polar interfaces, and these dipoles can vary between the non-polar and polar interfaces. The termination dependence of SBH has been observed at the Al/X junctions (X= Ge, GaAs, AlAs, and ZnSe) [14], the metal/MgO interfaces [15] and Si/$XO_2$ interfaces (X=Hf and Zr) [16]. Previous studies have also found an orientation dependence of SBH dipoles at the metal/$HfO_2$ interface [17,18].

Some exceptions to the MIGS model were found for semiconductors interfaces with metals which contain an underlying covalent lattice such as silicides. These interfaces apparently have weaker pinning, but actually have extra gap states which cause their SBHs to depend quite strongly on silicide work function [19,20].

Several problems can be broadly considered in terms of their metal/oxide interfaces. ZnO has well-known uses in optoelectronic devices, either by itself or alloyed with MgO [1,2,21], where SBHs affect injection efficiencies. MgO is a particularly important insulator for magnetic tunnel junctions (MTJs) [6]. An important factor of MTJs used in spin-transfer torque magnetic random-access memory (STT-MRAM) is that some properties such as their perpendicular magnetic anisotropy (PMA) could depend on the atomic nature of its magnetic metal-oxide interface and whether it has oxygen deficiency or off-stoichiometry. Thus, it is important to understand such relationships.

We thus consider four oxides (CdO, ZnO, MgO and SrO), covering various dielectric constant values. Although studies have been performed on some of these interfaces [15,22–26], to our knowledge there is no systematic theoretical work on the SBH trends with the interfacial structure, charge transfer and dipole layer for the different metal/oxide contacts. Therefore, we study different metals on these selected metal oxides, to investigate the microscopic mechanisms responsible for the chemical trends in SBHs.

## II. Calculation Method

The calculations use CASTEP code [27], with the plane-wave pseudopotential method and the Perdew-Burke-Ernzerhof version of the generalized gradient approximation (GGA-PBE) for the exchange-correlation functional. Norm-conserving pseudopotentials with a plane-wave cutoff energy of 680 eV are adopted. Pseudopotentials for oxygen are generated by the OPIUM method [28].

GGA functional has been widely used to calculate electronic structures. However, it is known to underestimate the band gap of semiconductors and insulators. In particular, ZnO and CdO are severely affected by this shortcoming, with the calculated band gap of ~0.8 eV for ZnO [29,30], while CdO even shows a negative indirect gap [31,32] (Table I).

The GW and similar methods [31] can correct these band structure errors, however, they are very time-consuming. The hybrid functional methods like the screened exchange (SX) [32,33] or Heyd-Scuseria-Ernerzhof (HSE) [34] functionals can also be used, they are more costly than GGA but not as much as GW. However, these methods can have difficulties with metallic systems and would be inefficient for some larger supercells used here for SBH supercell calculations.

We seek a lower-cost method to correct the band gap error to treat large SBH supercells. The GGA+U method was initially introduced for open-shell transition metals to describe the effect of electron-electron repulsion within the metal d states. This method was carried over to closed shell transition metal compounds with filled cation d states where a U potential on the d states would open up the band gap [29]. However, an unphysically large U value is often needed to give the correct band gap. Instead, a more moderate U interaction on both anion p states and cation d states of ZnO and CdO can be used to open up the gap, by correcting the overestimated p-d coupling [29,30], as shown below. In MgO and SrO, we show that a U potential can be applied to the O-2p states to widen their band gaps, without any d states being involved.

## III. Results and discussion

### A. Band Structures

The basic electronic structures of the oxides have a large impact on the SBHs, and the accurate description of band structures is the base of reasonable contact SBHs. To give a well-defined basis for the microscopic understanding of the interfacial electronic properties, it is required to correct the DFT-GGA error and reproduce the experimental values. The band structure of ZnO has been calculated by many workers [30–33,35,36]. The SX bands of ZnO [30,32,33,36] are essentially identical to the GW versions [31], whereas the HSE bands need the Hartree-Fock fraction α to be increased from 25% to 37% for this [35]. The band gap of ZnO can be corrected from its GGA form to agree with GW by adding U=4 eV to the Zn-3d states and U=9 eV to the O-2p states. With this reasonable combination of U values, the band gap is increased to 3.4 eV, as shown in the left panel of Fig. 1(b).

CdO has an unusual band structure, with an experimental valence band maximum (VBM) at L instead of Γ and a negative band gap in GGA. To fix the negative band gap problem, we include a U=2 eV on the Cd-4d valence states and U=4 eV on the O-2p states, right panel of Fig. 1(b). We obtain a lowest direct gap of 2.27 eV at Γ (compared to the experimental value of 2.28 eV), and an indirect gap of 1.17 eV, with VBM at L and a conduction band minimum (CBM) at Γ. Overall, the band structure is consistent with both theoretical [31–33,37,38] and experimental data [39,40]. The resulting band structures of ZnO and CdO are shown in Fig. 1, together with their GGA and SX band structures for comparison and their detailed band energies are summarized in Table I.

MgO has been the canonical cubic oxide insulator. It has the rock-salt crystal structure. It has a simple band structure with a p–like VBM at Γ and an s-like CBM at Γ [31]. It is often used as a substrate for catalysts and as an insulator in MTJs [6]. SrO is unusual in that while its valence band has the standard form with VBM at Γ, its conduction band has a local minimum at Γ but an absolute minimum at X.

The same principle of widening band gaps can be used for MgO and SrO, where O-2p states are used. We find that U=8.5 eV on O-2p states for MgO and U=8 eV on O-2p states for SrO can lower the VBM and give the correct band gap for each case. The calculated band gap for MgO is 7.79 eV, close to experiment [41]. For SrO, the direct band gaps at Γ and X are equal to 6.06 eV and 5.52 eV, respectively. The CBM of SrO is formed by Sr 5s states at Γ and by 5s and 4d states at X [42,43]. The lower CBM at X yields an indirect Γ-X gap of 5.36 eV, which is in good agreement with previous calculations [44]. Another notable feature of SrO is its narrow O 2p valence bandwidth which is due to its large ionicity. This bandwidth is found by other methods like GW, and by scaling in the tight-binding method by Pantelides [45]. The band structures of MgO and SrO are shown in Fig. 2, with band details summarized in Table I. Overall, the GGA+U bands are seen to be in good agreement with previous GW, SX and optical results [31–33,41,44].

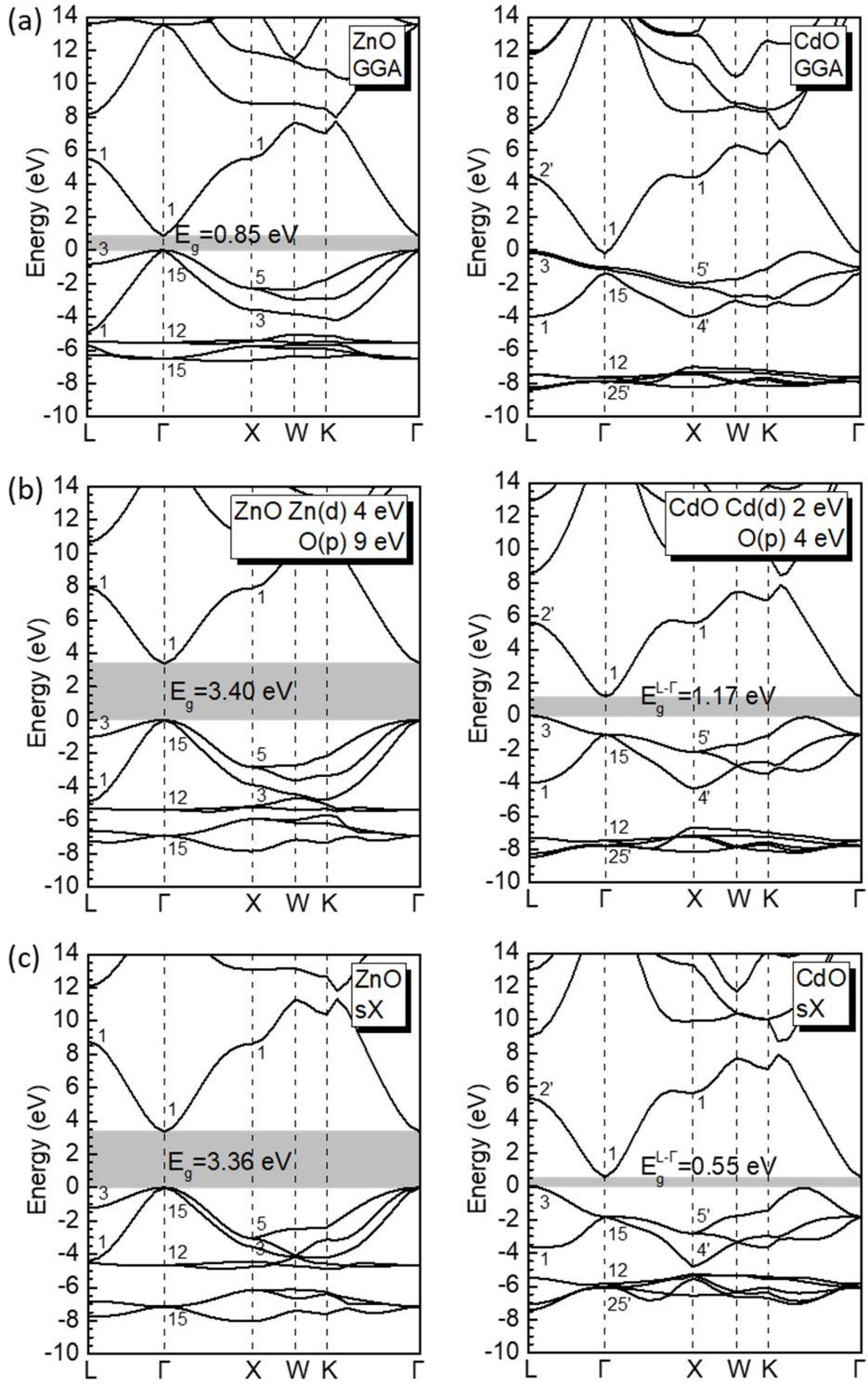

FIG. 1. Calculated electronic band structures of zincblende ZnO (left-panel) and rock-salt CdO (right-panel), using the (a) GGA, (b) GGA+U and (c) sX method, respectively. The minimum gap is labeled and shaded in grey. The energy zero is placed at the VBM.

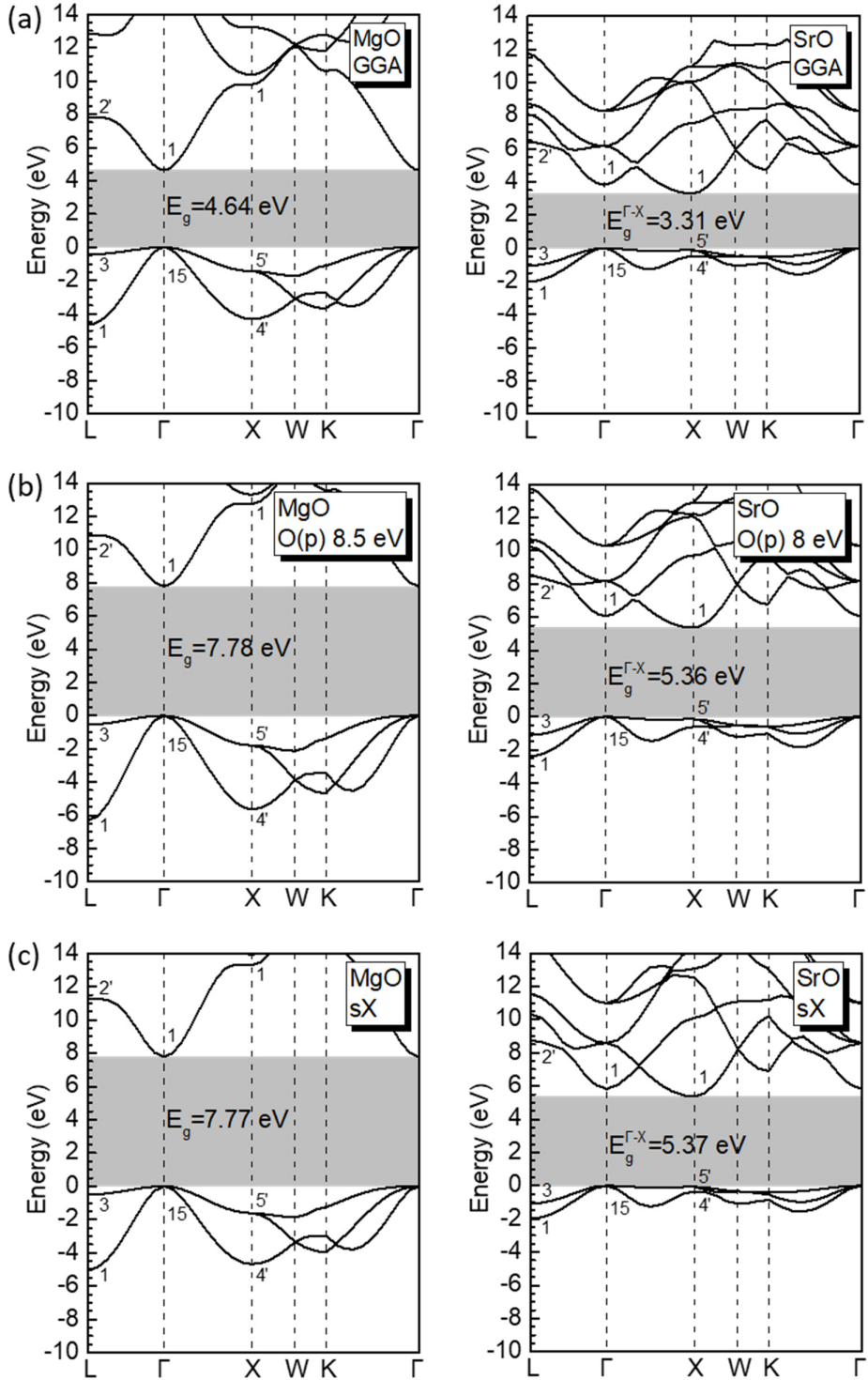

FIG. 2. Calculated electronic band structures of rock-salt MgO (left-panel) and SrO (right-panel), using the (a) GGA, (b) GGA+U and (c) sX method. The minimum gap is labeled and shaded in grey. The energy zero is placed at the VBM.

Table I. Band energies for cubic CdO, ZnO, MgO and SrO, calculated by GGA, GGA+U and sX methods, compared to experimental data. The position of the lowest conduction (c) bands and the highest valence (v) bands at high-symmetry points are studied. The uppermost VB at Γ is used as energy zero for ZnO, MgO and SrO. For CdO, energy zero is set at $L_{3v}$.

|  |  | $\Gamma_{1c}$ | $\Gamma_{15v}$ | $L_{2'c}$ | $L_{3v}$ | $X_{1c}$ | $X_{5'v}$ | Minimum gap |
|---|---|---|---|---|---|---|---|---|
| CdO | GGA | -0.15 | -1.4 | 4.47 | 0.00 | 4.37 | -2.07 | / |
|  | GGA+U | 1.17 | -1.10 | 5.62 | 0.00 | 5.58 | -2.12 | 1.17 (L-Γ) |
|  | sX | 0.55 | -1.85 | 5.32 | 0.00 | 5.65 | -2.82 | 0.55 (L-Γ) |
|  | Experiment | 2.28 [39], 2.32 (Γ-Γ) [40] | | / | | / | | 0.55 [39], 1.98 (L-Γ) [40] |
| MgO | GGA | 4.64 | 0.00 | 7.80 | -0.47 | 9.79 | -1.45 | 4.64 (Γ-Γ) |
|  | GGA+U | 7.78 | 0.00 | 10.86 | -0.51 | 12.73 | -1.79 | 7.78 (Γ-Γ) |
|  | sX | 7.77 | 0.00 | 11.29 | -0.45 | 13.29 | -1.60 | 7.77 (Γ-Γ) |
|  | Experiment | 7.77 (Γ-Γ) | | / | | / | | 7.77 (Γ-Γ) [41] |
| SrO | GGA | 3.81 | 0.00 | 6.44 | -1.02 | 3.31 | -0.16 | 3.31 (Γ-X) |
|  | GGA+U | 6.06 | 0.00 | 8.45 | -1.10 | 5.36 | -0.16 | 5.36 (Γ-X) |
|  | sX | 5.83 | 0.00 | 8.75 | -1.05 | 5.37 | -0.08 | 5.37 (Γ-X) |
|  | Experiment | 5.89 [42], 6.08 (Γ-Γ) [43] | | 5.97 [42] (L-L) | | 6.28 [42], 5.79 (X-X) [43] | | / |
|  |  | $\Gamma_{1c}$ | $\Gamma_{15v}$ | $L_{1c}$ | $L_{3v}$ | $X_{1c}$ | $X_{5v}$ | Minimum gap |
| ZnO | GGA | 0.85 | 0.00 | 5.48 | -0.83 | 5.52 | -2.32 | 0.85 |
|  | GGA+U | 3.40 | 0.00 | 7.90 | -1.02 | 7.94 | -2.81 | 3.40 (Γ-Γ) |
|  | sX | 3.36 | 0.00 | 8.71 | -1.28 | 8.63 | -3.10 | 3.36 (Γ-Γ) |
|  | Experiment | 3.44 (Γ-Γ) | | / | | / | | 3.44 (Γ-Γ) [1] |

## B. Electronic Structure at the Interface

We test the orientation dependence of the SBH in ZnO, MgO and SrO, which have wide direct band gaps, to ensure there is no interference between the oxide bands and the Fermi levels of the various metals which might affect overall trends. For CdO, the variation of SBHs is large compared to its small band gap, which could decrease the fitting accuracy, so we only focus on its non-polar interface. The non-polar (110) and polar (111) interfaces of ZnO are studied. For MgO and SrO, the calculations are based on the non-polar (001) and polar (111) interfaces. The effects of different polar terminations have already been treated [14,18]. So, in this work, we focus on the O-terminated interfaces for the polar case, because it is more energetically favorable for a metal contact.

The lattice mismatch and supercell periodicity between metals and oxides are listed in Table II. The sizable cell sizes needed for some cases explain why the faster GGA+U method is useful for some Schottky barrier calculations. Five layers of metal and seven layers of oxide are used. Since the metal work function is not so sensitive to the lattice constant, metals are strained to match the oxides' surface slabs if necessary. The interfaces are built using an interfacial supercell slab with a 15 Å vacuum slab. The geometries are relaxed to give energy difference below $5\times10^{-6}$ eV/atom and forces below 0.01 eV/Å. The SBH values are extracted using the core-level method [46,47] to increase the precision in determining the VBM, expressed by Eq. (4), where $E_{core}^{int}$ is the core level state in the interfacial model:

$$\phi_p = E_{core}^{int} + \Delta V - E_F \qquad (4)$$

Table II: Lattice matching in supercells of different metals and oxides. Taking non-polar CdO and Ru as an example, $\sqrt{8}\times\sqrt{8}=\sqrt{5}\times\sqrt{5}$ means that $\sqrt{8}\times\sqrt{8}$ sized supercell of Ru (001) surface is fitted with $\sqrt{5}\times\sqrt{5}$ sized supercell of CdO (001) surface.

| Oxide | Metal | Matching | |
|---|---|---|---|
| | | Non-polar | Polar |
| CdO | Ru, Os, Re, Rh, Ir, Pt | $\sqrt{8}\times\sqrt{8}=\sqrt{5}\times\sqrt{5}$ | / |
| | $MoO_3$ | $2\times2=\sqrt{5}\times\sqrt{5}$ | / |
| ZnO | Ti, Ru, Os, Re, Rh, Pd, Pt | $2\times2=\sqrt{3}\times\sqrt{3}$ | $2\times2=1\times1$ |
| | Ni | $\sqrt{7}\times\sqrt{7}=2\times2$ | $\sqrt{5}\times\sqrt{5}=1\times1$ |
| | $MoO_3$ | $\sqrt{3}\times\sqrt{3}=2\times2$ | $\sqrt{2}\times\sqrt{2}=1\times1$ |
| MgO | Sc, Hf, Zr, Ti, W, Ru, Re, Rh, Pd, Ir, Pt | $1\times1=1\times1$ | $1\times1=1\times1$ |
| | Cr | $\sqrt{5}\times\sqrt{5}=2\times2$ | $2\times2=\sqrt{3}\times\sqrt{3}$ |
| | $MoO_3$ | $\sqrt{5}\times\sqrt{5}=\sqrt{8}\times\sqrt{8}$ | $2\times2=\sqrt{7}\times\sqrt{7}$ |
| SrO | Sc, Hf, Zr, Ti | $\sqrt{5}\times\sqrt{5}=2\times2$ | $2\times2=\sqrt{3}\times\sqrt{3}$ |
| | Ru, Rh, Pd, Ir, Pt | $\sqrt{2}\times\sqrt{2}=1\times1$ | $\sqrt{7}\times\sqrt{7}=2\times2$ |
| | $MoO_3$ | $1\times1=1\times1$ | $1\times1=1\times1$ |

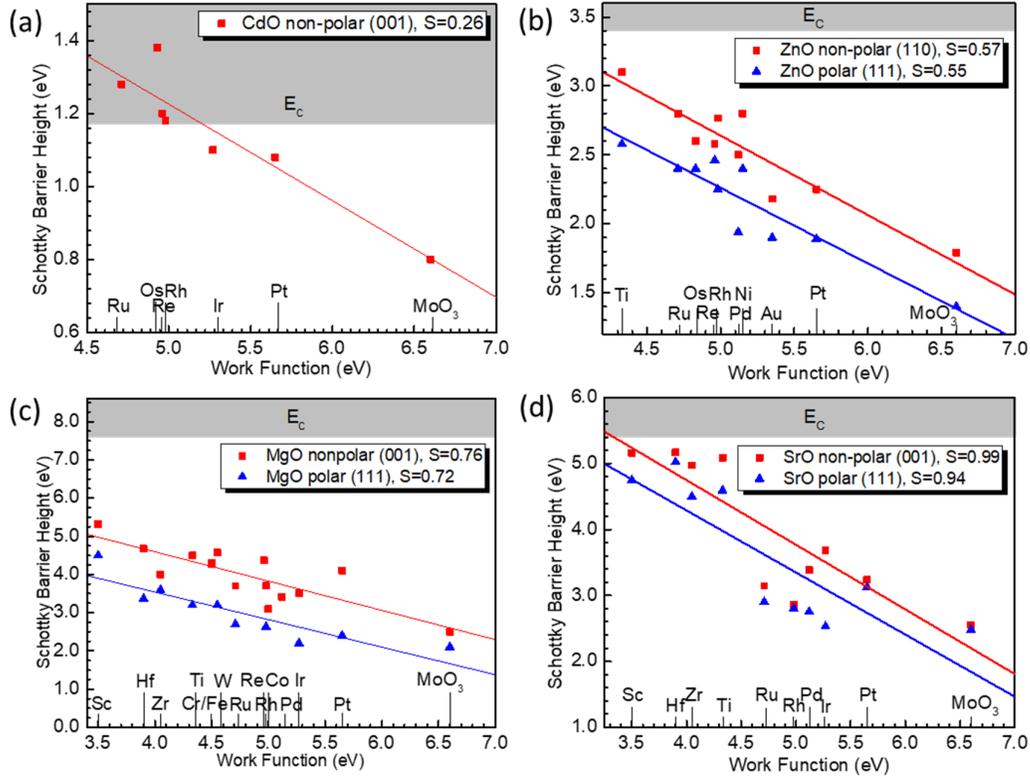

FIG. 3. Calculated p-type SBHs of various metals (a) CdO, (b) ZnO, (c) MgO and (d) SrO with different surfaces. Different orientations of one oxide are seen to have similar pinning factor S for ZnO, MgO and SrO. A clear shift of SBH is seen for different terminations. Work function values are from Michaelson [48].

Our work finds that for all four oxides, there is a linear dependence of SBH on metal work function, as shown in Fig. 3. A downward shift of the p-type SBHs can be seen for O-terminated polar faces from the non-polar ones, confirming that different terminations have a strong effect on the interface dipole [18].

We can calculate the pinning factor S from the fitted slope in Fig. 3. CdO has strongly pinned interfaces with S=0.26, while ZnO and MgO are marked by moderate S-parameters (~0.56 and ~0.74). SrO is weakly pinned and has S close to unity. Fig. 4 compares these calculated S factors with the experimental optical dielectric constants $\varepsilon_\infty$ [49,50], via the empirical MIGS model, Eq. (3). S follows a power law dependence on $\varepsilon_\infty$ for CdO, ZnO and MgO as predicted by the MIGS model. However, the metal/SrO contact shows weaker Fermi level pinning than this empirical trend. In this case, S is lower because of its large ionicity, as shown in Table III. The oxide ionicity increases from ZnO, CdO to MgO, and finally reaches the very high ionicity in SrO with nearly unity pinning factor.

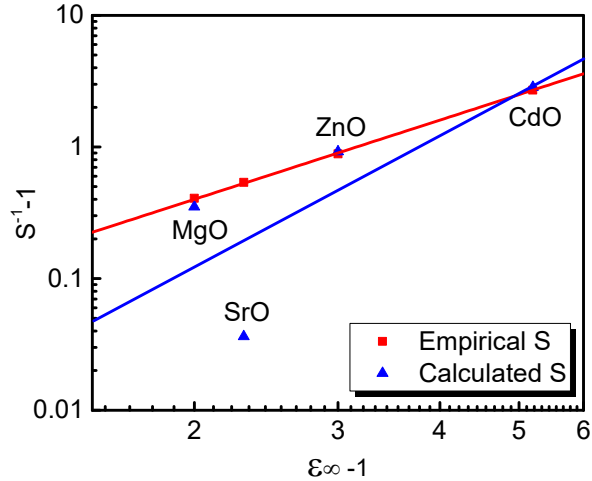

FIG. 4. Variation of calculated pinning factor S versus experimental optical dielectric constant $\varepsilon_\infty$, with the empirical formula for S from Eq. (3) plotted as a reference.

Table III. Pinning factor comparison between our calculated values from Fig. 3, and empirical values from Eq. (3), together with the difference between SBHs for polar and non-polar interfaces. The oxide ionicities are from Levine [50], and dielectric constants listed are experimental values [38].

|  |  | Calculated S | SBH shift (eV) | Ionicity | $\varepsilon_\infty$ | Empirical S |
|---|---|---|---|---|---|---|
| CdO | / | 0.26 | / | 0.78 | 6.2 | 0.27 |
| ZnO | Non-polar | 0.57 | 0.4 | 0.65 | 4.0 | 0.53 |
|  | Polar | 0.55 |  |  |  |  |
| MgO | Non-polar | 0.76 | 1.0 | 0.84 | 3.0 | 0.71 |
|  | Polar | 0.72 |  |  |  |  |
| SrO | Non-polar | 0.99 | 0.5 | 0.93 | 3.3 | 0.65 |
|  | Polar | 0.94 |  |  |  |  |

We now consider the interfacial electronic characteristics that cause pinning at the metal/oxide interfaces. We compare in Fig. 5 (left panel) the partial density of states (PDOS) for the metal/ZnO and metal/SrO interfaces, with the most prominent difference between their pinning strength. To demonstrate the effect of the metal on the interface oxide atoms, the PDOS of individual interface layers are plotted separately, together with the PDOS of the bulk compounds. We focus on the non-polar oxide interface, so the effect of additional interface dipole could be minimized. The modifications of the PDOS for the Pd/ZnO interface from the bare surface are quite pronounced. These interface states in the ZnO band gap can absorb charge transferred from the metal and effectively pin the Fermi level.

In contrast, SrO has much fewer induced gap states, leading to an almost unpinned interface. In both cases, the perturbation of the interface PDOS around the bulk oxide band gap consists of MIGS, and their extent is determined by the energy of the Pd states. For Pd/ZnO, the MIGS extends as far as the third layer of ZnO from the interface, while the MIGS at Pd/SrO interface extends only to the first SrO layer. Fig. 5 also shows the modification of PDOS on the metal side, and the metal does not simply screen out the influence of the semiconductor.

Although MIGS are seen at these interfaces, especially for ZnO, it is believed that the pinning effect of MIGS is low and much of the experimentally observed pinning is caused by defects [2]. In the case of ZnO, the O vacancy has been shown to provide pinning [51], at an energy of 0.7 eV below the conduction band edge. Taking these into consideration, the MIGS theory still provides a successful prediction for the interface pinning effect.

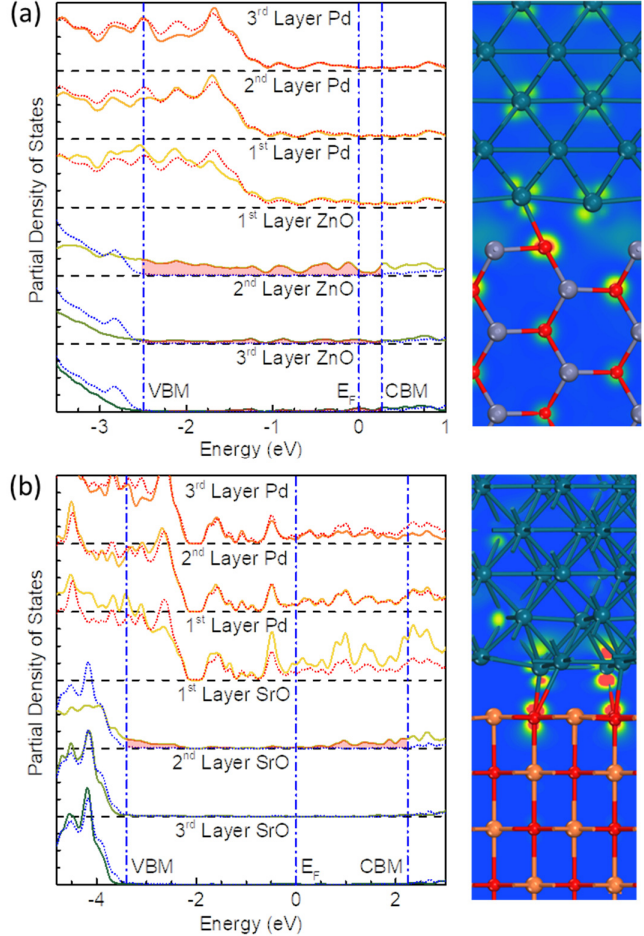

FIG. 5. Partial DOS of (a)ZnO (b)SrO individual layers from the interface model, with VBM, CBM and Fermi level labeled. Pd is the contact metal. Interface induced gap states are shaded in red. The dotted red lines indicate the PDOS of metal Pd and the dotted blue lines indicate the PDOS of ZnO or SrO. The right panel for each image shows the wavefunction with energy around the metal Fermi energy.

Another interesting aspect of our results is that, although different faces of the same semiconductor share the same pinning strength S at the interface, there is a great difference between the non-polar and polar terminated interfaces. The SBHs of ZnO vary by roughly 0.4 eV higher when the orientation is changed from polar (111) to non-polar (110), while in the cases of MgO and SrO, the (001) SBH values lie approximately 0.5 eV and 1 eV above the O-terminated (111) SBHs. It is notable that the shift in SBH changes inversely to the oxides' dielectric constant, which corroborates the results of previous studies [14,18]. The dependence of the SBH on the interface stoichiometry is beyond the description of MIGS model. This effect can be explained by the additional dipole created at the polar interface, which consists of the charge transfer effect and the image effect. Firstly, the charge transfer between metal and the induced interface states can bring in a net charge at the interface, whose size depends on the Fermi energy compared to the charge neutrality level (CNL) energy. The charge density difference $\Delta\rho$, is calculated in Fig. 6 as follows:

$$\Delta\rho = \rho_{metal/oxide} - \rho_{metal} - \rho_{oxide} \quad (5)$$

where $\rho_{metal/oxide}$ is the total charge density of the metal/oxide contact. $\rho_{metal}$ and $\rho_{oxide}$ are the charge density of isolated metal and oxide slabs with the same geometries as in the combined system.

Fig. 6 shows the 2D energy-sliced charge redistribution at different interfaces. The metal/ZnO and metal/MgO contacts are chosen here as their SBH shifts are significantly different. From the oxide layers near the interface, it is clear that only oxygen atoms are responsible for the charge transfer with the Pd atoms. The charge accumulation around O and Pd has the characteristics of an s-d hybridized orbital and occupies charge accumulation, while a clear depopulation of the O-p orbital and Pd-dz2 orbital can be observed. It is also worthwhile noting that, the oxygen atoms show greater charge loss and a smaller region of remaining charge density over the non-polar interface region compared to the polar case. The enhanced chemical interaction and charge transfer at the polar interfaces result in the Schottky barrier conversion.

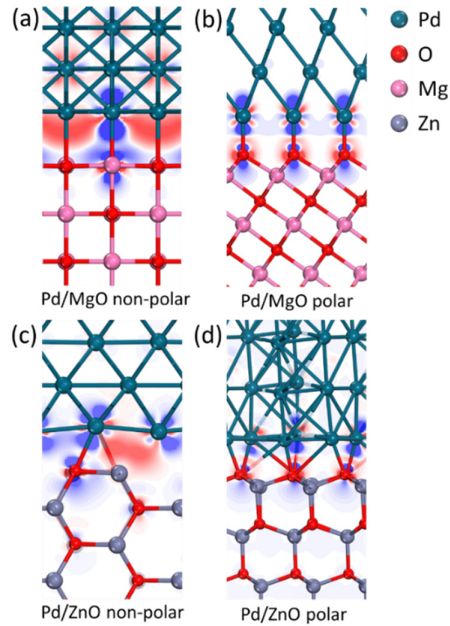

FIG. 6. Charge density differences compared for (a) Pd/MgO non-polar, (b) Pd/MgO polar, (c) Pd/ZnO non-polar and (d) Pd/ZnO polar interface. Red indicates regions of charge accumulation and blue indicates regions of charge depletion.

It is finally interesting to compare models for the metal/MgO interfaces in MTJs especially given their recent technological significance. MgO acts as the tunnel barrier between two electrodes of a body-centered cubic (bcc) Fe/Co alloy, to which boron is initially added as an amorphizing agent [6]. Annealing then releases boron into the underlying Ta layer and causes Fe/Co to crystallize by templating on the cubic MgO. The final interface is surprisingly flat and well-separated [6,52]. The tunneling between the electrodes occurs by the MIGS and is central to the MTJ's operation. The tunneling probability depends sensitively on the MIGS symmetry [53] and is degraded by off-stoichiometry at the metal/MgO interface [54]. Stability of spin polarization favors the electrode spin directions to lie perpendicular to the MgO plane [55]. There have been two models of this process, one dependent on an off-stoichiometric layer between MgO and Fe [56], and a second due to an interaction between the O-$2p_z$ state and the interfacial Co-$d_{z2}$ state which aligns the minor spin direction [57,58], the perpendicular magnetic anisotropy (PMA). Recent high-resolution TEM imaging verifies the sharpness of the Co,Fe|MgO interface and suggests that any non-stoichiometry interfacial layer is low [52]. This shows the importance of these metal-insulator interfaces and SBHs for MTJs.

## IV. Conclusions

In conclusion, SBHs of different metal/oxide interfaces are systematically studied. Different orientations of the same oxide are found to have similar pinning strength S, which increases with the oxide ionicity. For more covalent oxide like CdO, MIGS is the main contributing factor of Fermi level pinning. For the more ionic ones like ZnO and MgO, although MIGS are a less important cause of pinning than defects experimentally, the pinning factor predicted by first-principles calculation still gives the same value as that found by the empirical model (Eq. (3)). SrO is considered to be unpinned and the simple model is not applicable due to its high ionicity. We can conclude that MIGS could be a good simplified model, and its prediction is more accurate for less ionic oxides. Despite sharing the same pinning slope, there is a significant upward shift of SBH values when the orientation of oxide is changed from polar to non-polar. The difference in SBHs could be explained by the charge transfer at the interface induced gap states and the image charge effect. The polar terminations have stronger charge redistribution over the interface, which gives rise to the formation of additional interface dipole and raise the oxide band with respect to the metal. The magnitude of shifting is found to vary inversely with the oxide dielectric constant.

The authors acknowledge funding is from EPSRC Grant No. EP/P005152/1. We acknowledge the support from Supercomputing Wales under the project SCW1070.